# A quasi-commercial Terahertz detector based on MIR thermopiles

F. Voltolina[1,*], U. Dillner[2], E. Kessler[2], P. Haring Bolivar[1]

[1]*Institute of High Frequency and Quantum Electronics, University of Siegen, Hölderlinstr. 3, D-57076 Siegen, Germany*

[2]*Institut für Photonische Technologien, Albert-Einstein-Straße 9, 07745 Jena, Germany*

[*]*Corresponding Author:* francesco.voltolina@ieee.org

**Abstract:** Integrating a thermopile with a low power operational amplifier is an effective and cost-efficient approach to obtain a high performance compact Terahertz power detector. Here we present the development of such an integrated detector, including the fundamental building blocks, the final realization and related characterization. The responsivity and bandwidth of the detector are measured in a standard THz CW setup consisting of a closed cycle helium cryostat and a Quantum Cascade Laser (QCL). The predicted results are compared with measured data, showing a good agreement. Advantages of existing Golay cell THz detectors in both sensitivity and NEP figure vanish, in practical use, due to the enormous increment in dynamic range provided by the linear operation of the thermopile up to 100 mW of incident power while offering a NEP below 1nW/√Hz and a -3dB bandwidth of 6,8 Hz.

## Introduction

The thermopile is, historically, the first practical infrared detector providing an electrical output signal: its invention and initial use dates back to 1820 with the physicists Oersted, Fourier, Nobili and Melloni. Despite its current availability as a mass produced device, often embedded in low cost consumer applications, to the best of our knowledge it has received only marginal attention in the THz community.

While in the mid-infrared (MIR) it is widely used in applications ranging from remote sensing thermometers to satellite attitude sensors, it is typically limited to the role of wide area power meter detector in the THz range, with only one recent exception known to the authors [1, 2].

Here, we investigate a commercially available thermopile, the TPS334 (see Fig. 1) from Perkin Elmer [3], and a custom thermopile from the Institute of Physical High Technology e.V. (IPHT) of Jena [4], both designed for the MIR, to determine their suitability as a flexible and compact THz detector.

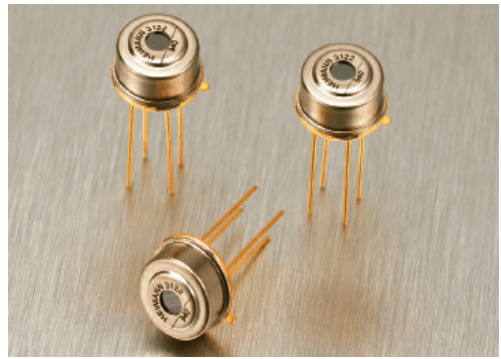

Fig. 1 Picture of the Perkin Elmer commercial thermopiles (c) PE

**Device Integration**

The TPS334 single pixel detector is characterized by a sensitive area of 0.7x0.7 mm$^2$ and responsivity of 55 V/W with a NEP of 0.64 nW/√Hz, for a 500K MIR blackbody radiator at 1 Hz chopper frequency, without window. The -3dB point in the frequency response is reached at 11 Hz, making its use attractive for fast power detection or even for a THz sensing or imaging application with a a limited amount of sensing sites or image pixels. One potential application is e.g. in a DNA biochip reader system [5].

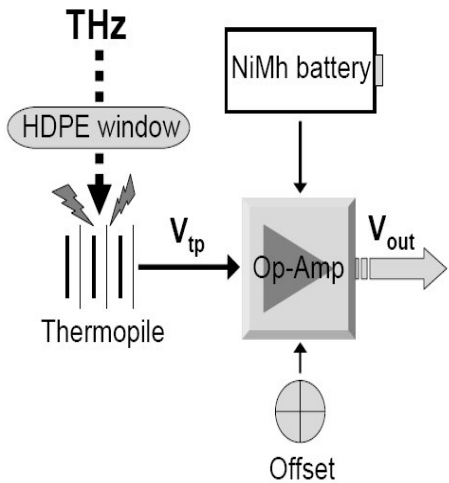

Fig. 2  Working principle of our compact integrated detector

The custom IPHT-Jena single pixel detector is characterized by a circular sensitive area of 1.0 mm$^2$ and responsivity (at MIR) higher than 47 V/W, again without window. To maximize the detector flexibility we realized it as a compact detector integrating the thermopile chip and a battery operated low power amplifier chip in one compact enclosure, as seen schematically in Fig. 2.

The detector system consists of a high stability aluminum block containing all the components and sealed with iron loaded epoxy resin for improved thermal and mechanical stability, as pictured in Fig. 3.

After initial test we integrate two final versions of the detector using either of the two alternative thermopile chips, and a common amplifier stage based on the Analog Device AD8538, a very high precision amplifier featuring extremely low offset voltage, low input bias current, and low power consumption: ideal for battery operation. Typical battery consumption is expected to allow several weeks of continuous measuring before the small NiMh battery needs to be recharged (given the minimal 250 µA max. power consumption and the 80 mAh battery).

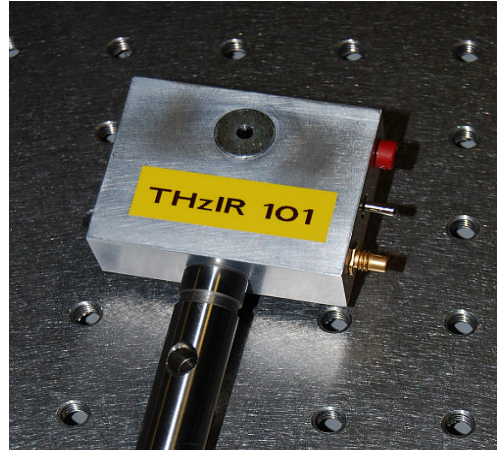

Fig. 3  Picture of a pre-series detector we have built

**Experimental setup**

The characterization setup uses a 2.5 THz QCL laser [6] cooled by a closed cycle Helium cryostat (model CCS-100/204 from Janis Research), a single high resistivity (HR) silicon lens (Focal length = Diameter = 25 mm.) to focus the radiation on the detector of choice and a Newport 2-axes high resolution motor stage for alignment (see Fig. 4).

A Golay cell and a Thomas Keating photo acoustic THz detector are used in order to allow a systematic detector inter-comparison and calibration.

The latter detectors require a lock-in amplifier based detection in order to reject background noise and artifacts, stemming from their broadband response reaching into the strong MIR background, and especially because they are AC coupled. To ensure inter-comparability, we initially test our detector and compare it with the other detectors using lock-in detection at several frequencies in order to characterize the modulation response and to find the -3dB point of both thermopiles. Additionally, since our detectors are DC coupled, we alternatively evaluated their performance connecting them directly to a standard ADC system.

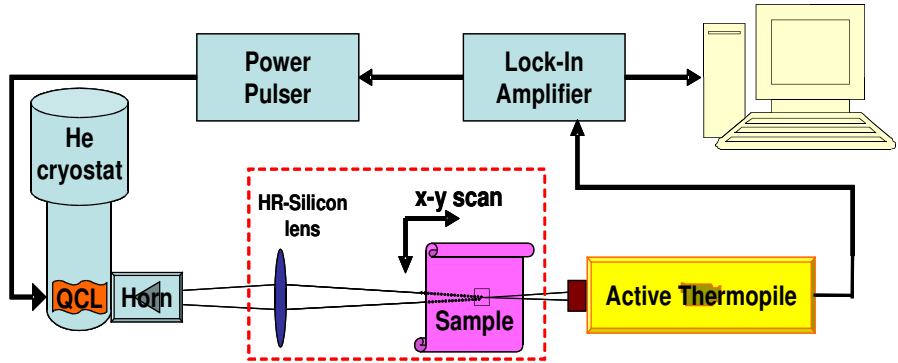

Fig. 4 Schematic drawing of our experimental setup

**Results**

The provided measurements setup has immediately shown that to achieve good readings is as simple as to find the correct positioning of a single lens: an HR silicon element that concentrates the radiation on the chosen detector. Our custom floating suspension for the cryostat is very effective in absorbing vibration issues so that even the sensitive photo acoustic detectors can operate optimally. Additionally, a temperature controlled and screened environment minimizes the systematic errors.

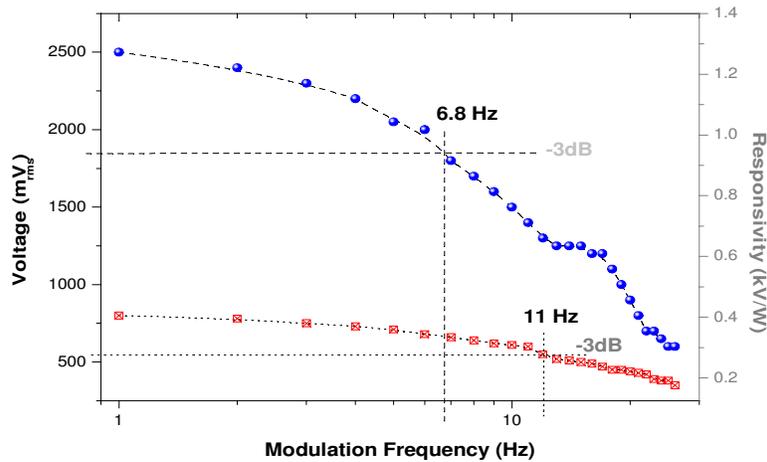

Fig. 5 Voltage output and responsivity versus modulation frequency graph for two different thermopile chips illuminated with a 2.5 THz QCL (at 2.0 mW average Power).

Blue circles = IPHT-Jena thermopile chip; Red squares = TPS334 PE thermopile chip

The advantages of a reference Golay cell photo acoustic detector in both sensitivity and NEP values vanishes in practical use due to the enormous increment in dynamic range provided by the linear operation of the thermopile up to 100 mW of incident power. No need of absorbers to limit the power to a tolerable level for the detector (damage power for a Golay detector is, on the contrary, in the range of only 20 μW) means all power of the source is available to increase the dynamic range of the measurement.

At the power level available from the modern QCL sources (in the order of tenths of milliwatts), the system SNR with our custom detectors is better than 50dB without the use of a lock-in amplifier. Higher SNR can be obviously achieved with a lock-in amplifier, which can be additionally tuned to the cryostat 1st order vibrations in order to eliminate, by averaging, artefacts due to standing waves effects, when acquiring images scanning those detectors across some samples in presence of strongly interfering stationary waves.

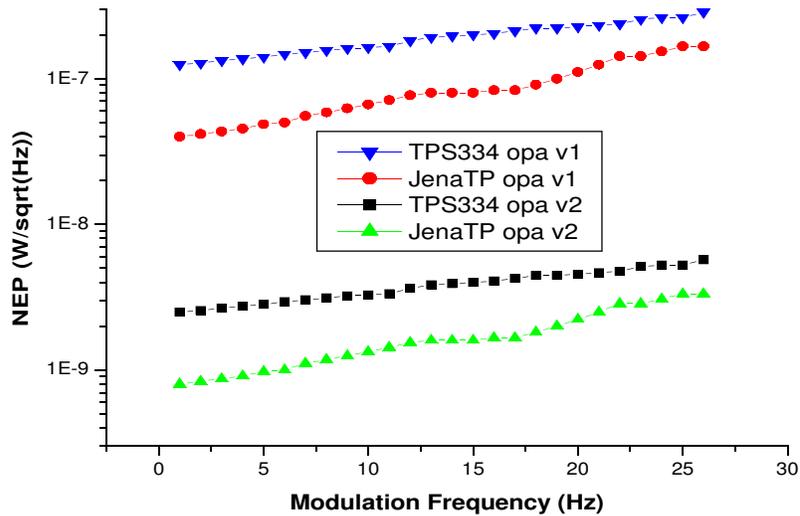

Fig. 6 Noise equivalent power (NEP) versus modulation frequency graph for two different thermopile chips and two different configurations for their operational amplifier

Comparison between calibrated and uncalibrated detectors has been used to validate the given MIR power response figures of our thermopiles and to measure the response of our custom detectors at THz frequencies. The voltage responses in dependence of the modulation frequency are available in Figure 5. From these values, using the figures for the power available from our laser outside the cryostat, after the SI lens, we were able to calculate the system responsivity of the two thermopile detectors, including the amplifier, to be 1 kV/W and 0.3 kV/W for respectively the IPHT-Jena and the Perkin Elmer based detector. Great figures either from the convenience point of view, of providing a high responsivity, and more importantly from the practical, as the thermopile detectors do not exhibit saturation at the readily available high THz power levels of QCL.

The comparison of those values to the given MIR responsivity, while factoring in the amplifier that has a gain of 1000, shows that the infrared absorber used in these devices manifest a definite responsivity loss.

In the THz region they are exhibiting a sensitivity loss of around 50 and 150 mainly due to absorber losses. At those longer wavelengths the IPHT-Jena based detector is coping better than the Perkin Elmer one and this is related to the special absorber coating technology used in the former device. Nonetheless, after calibration, they are showing perfect conversion consistency in the 100 µm region available with the QCL.
Additional tests with 2 electronic sources at 600 GHz and at 140 GHz show that substantially greater losses in efficiency start only in the region where the pixel dimension approaches the wavelength of the incident radiation used for the measurements. This is a positive aspect that avoids using multiple calibration factors for different frequencies.

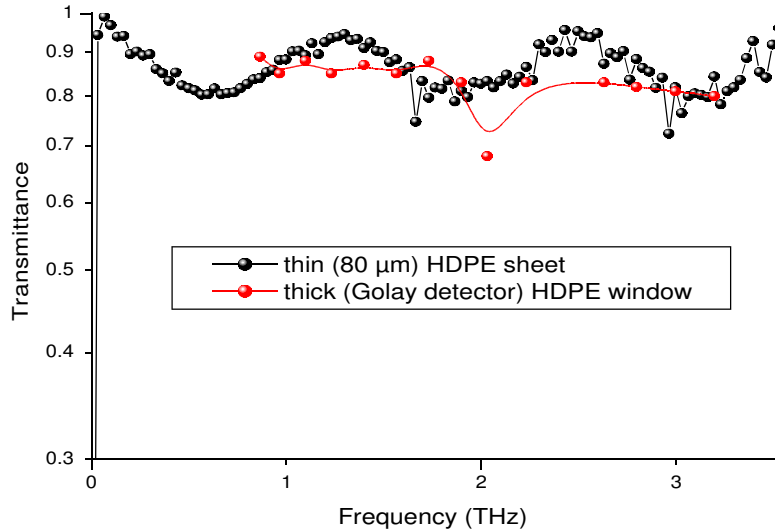

Fig. 7 Transmitted Power over Frequency for our sample of high density polyethylene window, compared to that given for the HDPE Golay detector window

Furthermore, we calculated from the measured responsivity and noise at the output of the amplifiers an additional graph showing the noise equivalent power versus the modulation frequency, as presented in Fig. 6. For our custom detectors we used a window material to protect them from oxidation and provide a proper nitrogen atmosphere for their operation using a thin (80 µm) HDPE sheet. A broadband measurement of its transmittance, taken in a standard TDS THz setup, is shown in Figure 7.
This graph shows a nice flat behaviour for the chosen window and it can be used to normalize the response at different THz frequencies.
Additionally, in the same graph, it is possible to compare our measurements with that presented in the Golay detector manual for its thick HDPE window [7].

**Conclusions**

A complete imaging system has been prepared that uses the 2.5 THz QCL laser cooled by a closed cycle Helium cryostat (model CCS-100/204 from Janis Research), the single high resistivity silicon lens (Focal length = 25 mm.) and a Newport 2-axes high resolution motor stage.

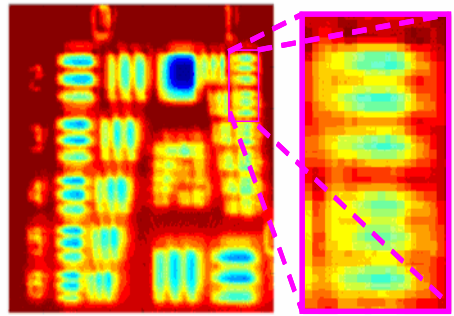

Fig. 8 Standard USAF resolution target: 2 and 2.24 line pair/mm

A simple test phantom was prepared, using a mail envelope filled with different materials known to cover a broad absorption range at THz frequencies, and imaged (see Fig. 9). Images taken at 100µm pixel step size drastically outperform that acquired with a Golay detector and according to transmission imaging of another sample (a standard USAF resolution target, chrome on fused silica) even 250µm features are well resolved (see Fig. 8).

The advantages of the Golay cell in both sensitivity and NEP values vanishes in practical use due to the enormous increment in dynamic range provided by the linear operation of the thermopile up to 100 mW of incident power.

The results have demonstrated the ability of thermopile devices to perform as practical and sensitive power detectors for THz radiation.

An ample margin for improvement can be expected by dedicated absorber developments.

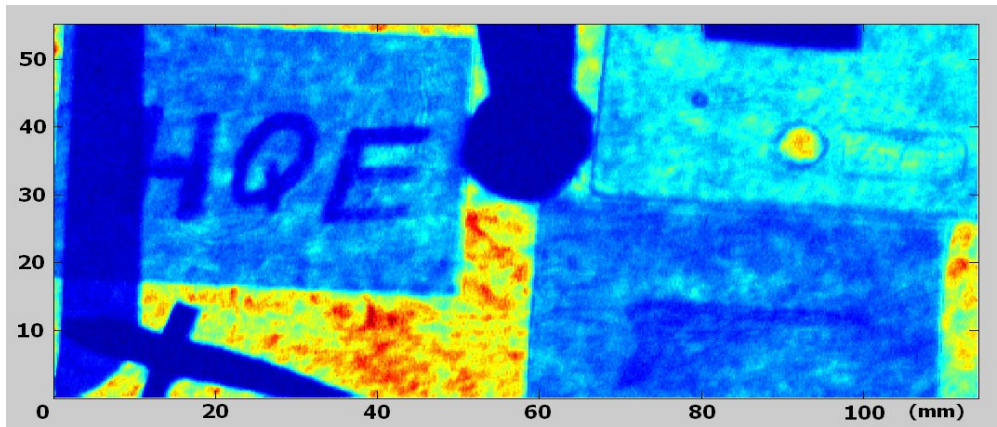

Fig. 9  Transmission image of an envelope at 2.5 THz.

**Acknowledgments**

This work was financially supported by the European Framework VI Project Teranova *(IST-511415)*.